# Rectangular carbon nitrides C$_4$N monolayers with a zigzag buckled structure: Quasi-one-dimensional Dirac nodal lines and topological flat edge states

Linyang Li,[a] Jialei Li,[a] Yawei Yu,[a] Yuxuan Song,[a] Jia Li,[a] Xiaobiao Liu,[b] François M. Peeters,[c,d] Xin Chen,[e,*] Guodong Liu[a,f*]

[a]*School of Science, Hebei University of Technology, Tianjin 300401, China*

[b]*School of Science, Henan Agricultural University, Zhengzhou 450002, China*

[c]*Centre for Quantum Metamaterials, HSE University, 101000 Moscow, Russia*

[d]*Department of Physics, University of Antwerp, Groenenborgerlaan 171, B-2020 Antwerp, Belgium*

[e]*Thermal Science Research Center, Shandong Institute of Advanced Technology, Jinan 250103, China*

[f]*State Key Laboratory of Reliability and Intelligence of Electrical Equipment, and School of Materials Science and Engineering, Hebei University of Technology, Tianjin 300130, China*

*Corresponding authors.

E-mail address: xin.chen@iat.cn (X. Chen), gdliu1978@126.com (G. Liu)

**Abstract**

Due to the flexibility of C and N atoms in forming different types of bonds, the prediction of new two-dimensional (2D) carbon nitrides is a hot topic in the field of carbon-based materials. Using first-principles calculations, we propose two $C_4N$ monolayers with a zigzag buckled (ZB) structure. The ZB $C_4N$ monolayers contain raised-C (raised-N) atoms with $sp^3$ hybridization, different from the traditional 2D graphene-like carbon nitride materials with $sp^2$ hybridization. Interestingly, the band structures of the ZB $C_4N$ monolayers exhibit quasi-one-dimensional (quasi-1D) Dirac nodal line that results from the corresponding quasi-1D structure of the zigzag carbon chains, which is essentially different from the more common ring-shaped nodal line. The quasi-1D Dirac nodal line exhibits the following features: (i) gapless Dirac points, (ii) varying Fermi velocity, and (iii) slightly curved band along the high-symmetry path. All these features are successfully explained by our proposed tight-binding model that includes interactions up to the third nearest-neighbor. The Fermi velocity of the 2D system can reach $10^5$ m/s, which is promising for applications in high-speed electronic devices. The topological flat band structure determined by the Zak phase and band inversion of the corresponding 1D system is edge-dependent, which is corresponding to the Su-Schrieffer-Heeger model, providing to rich physical phenomena.

# 1. Introduction

When electrons are limited to a one-dimensional (1D) area, many important physical phenomena may emerge. Recently, Si nanoribbons composed of pentagonal Si rings have been achieved and the corresponding 1D Dirac Fermions with a high Fermi velocity ($1.3\times10^6$ m/s) were confirmed by experimental measurements and theorical calculations [1]. These Si nanoribbons were found to be complete 1D system, which is different from the layered van der Waals (vdW) material $NbSi_xTe_2$ that shows a quasi-1D behavior in its band structure [2]. By tuning the value of *x*, the Dirac Fermions can live in different dimensions, which depends on the interaction between the $NbTe_2$ chains [3,4]. Since the interaction between the chains is weak, and therefore cannot affect the main Dirac band structure, and therefore it should be regarded as a quasi-1D Dirac nodal line. Hence, the three-dimensional (3D) systems $NbSi_xTe_2$ offer a good platform to study such a quasi-1D physics that may be relevant for high-speed electronic device applications. A question arises: Can we find quasi-1D band structure in two-dimensional (2D) systems? In this work, we will show that in 2D carbon nitrides this is possible.

Two-dimensional carbon nitrides have received a lot of attention due to their outstanding physical and chemical properties. Due to the great flexibility of C and N atoms, many different 2D carbon nitrides were proposed, which are indicated by the chemical formula $C_xN_y$, and are an important class of 2D materials beyond graphene [5]. The 2D carbon nitrides are well known to exhibit strong and stable components, owing to the formation of covalent bonds between the C-C/C-N atoms. In theory, the carbon nitride monolayers can not only exhibit promising mechanical, thermal, electronic, magnetic, and optical properties, but also have been extensively proposed as a possible candidate for many applications, such as catalysis, gas purification, and energy storage, including Li/Na/K-ion battery, Li-S battery, solar Cell, and hydrogen storage [6]. Experimentally, many 2D carbon nitrides have been achieved, including $C_3N_4$ [7],

C$_2$N [8], C$_3$N [9], C$_3$N$_3$ [10], and C$_6$N$_7$ [11]. Due to the progress in theoretical design and experimental fabrication, a large family of 2D carbon nitrides is appearing at the horizon. Among the family of 2D carbon nitrides, the C$_3$N$_4$ monolayer is the most popular member. It with few layers has a large specific area for reactive sites and intimate interface contact, and therefore allows for more functional modifications that are important for target-specific applications [12]. The C$_3$N$_4$ monolayer has two important kinds of N atoms: edge-N and graphitic-N [13]. The edge-N atom bonds to two C atoms, leading to a nanoporous structure, such as C$_2$N [14], C$_3$N$_2$ [15], C$_3$N$_3$ [16], C$_4$N$_4$ [17], C$_7$N$_3$ [18,19], C$_{10}$N$_3$ [20], and C$_{19}$N$_3$ [21]. The graphitic-N atom bonds to three C atoms, leading to a graphene-like structure, such as C$_3$N [22], C$_4$N [23], C$_5$N [24-26], and C$_6$N$_2$ [27,28], which are similar to monolayers of carbon allotrope and not limited to the six-ring of C-N/C atoms. More complex carbon nitride monolayers can be designed by combining edge-N and graphitic-N, including C$_5$N$_2$ [29], C$_5$N$_4$ [30], C$_7$N$_6$ [31,32], C$_9$N$_4$ [33,34], and C$_9$N$_{10}$ [35], which are the important basis for the family of 2D carbon nitrides. However, another kind of N atoms should also lead to a stable 2D structure, the raised-N atoms. Those monolayers with raised-N atoms are very rare [36-38], and their structures and properties are poorly explored.

There are two ways of constructing the monolayer with the raised-N atoms: (i) adding N atoms on graphene, and (ii) adding C atoms on 2D planar structure of carbon nitride. Taking the dumbbell (DB) C$_4$N as an example [36], the two C$_4$N monolayers can be obtained by adding C atoms on the hexagonal C$_3$N (h-C$_3$N) monolayer [9,39]. Recently, Guo *et al.* proposed three new C$_3$N allotropes with a rectangular lattice [40], which are C$_3$N-S1, C$_3$N-S2 and C$_3$N-S3. All the C/N atoms in these structures have $sp^2$ hybridization, similar to the experimental h-C$_3$N monolayer. Since $sp^3$ hybridization is also possible for the C/N atoms, and can be realized in the C$_3$N-S2 monolayer by adding C atoms forming two new C$_4$N monolayers with a rectangular lattice. Because one C atom can bond at most four C/N atoms, the C$_3$N-S1/C$_3$N-S3 monolayer cannot lead to a reasonable structure with raised-N atoms. In this work, we predict two new C$_4$N monolayers from the C$_3$N-S2 monolayer, and calculate their structure, stability, and electronic band structure using first-principles calculations. Surprisingly, a quasi-1D Dirac nodal line is found in the energy spectrum, which is essentially different from the more common ring-shaped nodal line, suggesting a realistic material platform for exploring quasi-1D Dirac fermions in 2D systems [41].

## 2. Computational method

The first-principles calculations were performed using the Vienna *ab-initio* Simulation Package (VASP) code, implying density functional theory (DFT) [42-44]. The electron exchange-correlation potential is treated in the form of Perdew-Burker-Ernzerhof (PBE) functional in the generalized gradient approximation (GGA) [45]. A plane-wave basis set with a cut-off energy of 520 eV was used. The Brillouin zone (BZ) is sampled by a K-mesh grid of 23×27×1. The convergence criteria for the total energy of 10$^{-5}$ eV for structural optimization and 10$^{-7}$ eV for other calculations. The convergence standard for force was set to 0.01 eV/Å. The vacuum space was set to at least 20 Å along the vertical direction to avoid interaction between adjacent layers. The phonon spectra were calculated using a supercell approach within the PHONOPY code to evaluate the dynamical stability [46].

## 3. Results and discussion
### 3.1 Structural property

Before discussing the new C$_4$N monolayers, we focus on the C$_3$N-S2 monolayer. A top/side view of this monolayer is presented in Fig. 1(a). It is a planar structure and exhibits a rectangular lattice with the space group *Pmma* (51). The optimized lattice constants are *a* = 4.864 Å and *b* = 4.204 Å, which are in good agreement with previous DFT calculations [40]. Lines through the N atoms along the *x*-axis have a zigzag shape, and thus this $C_3N$-S2 monolayer can be regarded as a zigzag planar (ZP) $C_3N$ monolayer. By adding a C atom on the N atom of the ZP $C_3N$ monolayer, two buckled monolayers of $C_4N$ can be obtained, as shown in Fig. 1(b) and 1(c). There are two possibilities for the buckled structures: (i) the raised-N/raised-C (N/$C_R$) atoms are all on the same side of the monolayer along the *z*-axis, and (ii) the N/$C_R$ atoms alternate on the two sides of the monolayer along the *z*-axis. The unit cell of $C_4N$ monolayer includes eight C atoms (six $C_P$ atoms and two $C_R$ atoms) and two N atoms. It is obvious that these two monolayers have a similar geometric structure, and the main difference is the vertical position of the N/$C_R$ atoms. For the two buckled $C_4N$ monolayers, lines through the N/$C_R$ atoms along the *x*-axis also form a zigzag shape, which we call zigzag buckled (ZB) $C_4N$-I and ZB $C_4N$-II, respectively. The ZB $C_4N$-I monolayer is a rectangular lattice with the space group *Pma*2 (28) while the ZB $C_4N$-II monolayer is a rectangular lattice with the space group $P2\_1/m$ (11). To better understand the difference between the three monolayers (ZP $C_3N$, ZB $C_4N$-I, and ZB $C_4N$-II), we will compare the structures from their structural symmetry, bond length, and orbital hybridization.

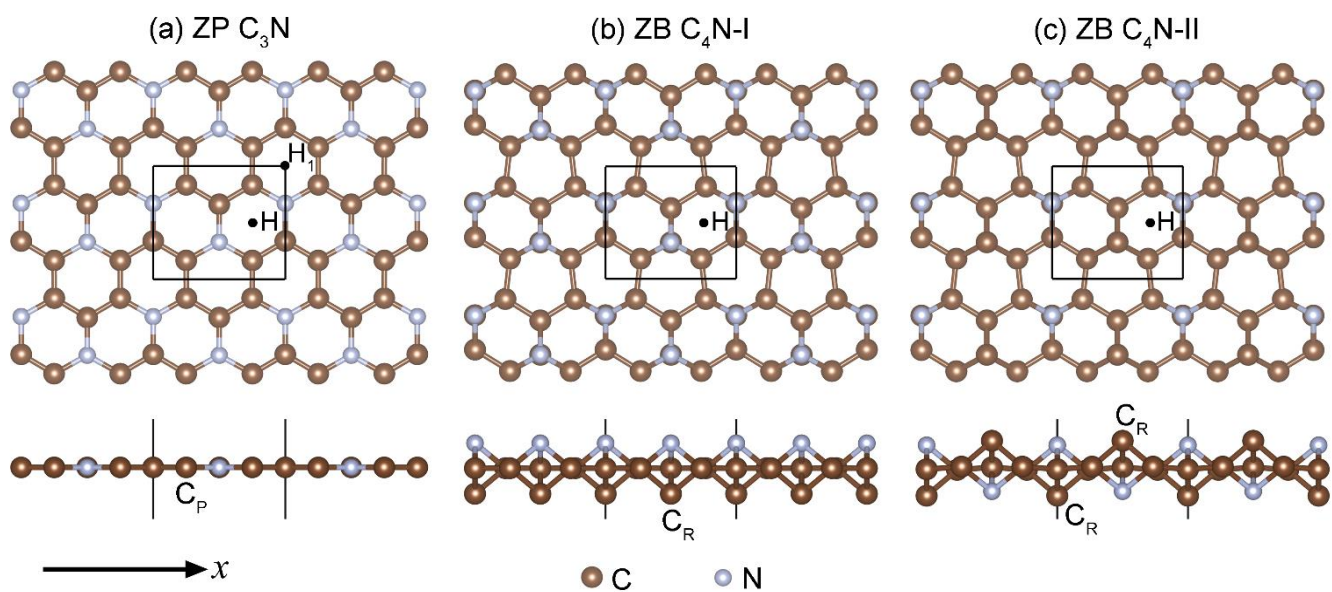

**Fig. 1.** Top and side views of ZP $C_3N$ monolayer (a), ZB $C_4N$-I monolayer (b), and ZB $C_4N$-II monolayer (c). The brown balls are the C atoms and the grey balls are the N atoms. The raised-C atoms are labeled as $C_R$ and the planar-C atoms are labeled as $C_P$.

As shown in Fig. 1, the main symmetry is the glide mirror symmetry for the three monolayers, and the symmetry mirror planes are perpendicular to the *x*-axis. In the ZP $C_3N$ monolayer, there are two kinds of six-rings (C4-N2 and C5-N1), where the centers are labeled as H and $H_1$, respectively. The H point is a center of inversion symmetry and the line parallel to the *z*-axis through the H point is a $C_2$ rotation symmetry axis. In the ZB $C_4N$-I monolayer, the $C_2$ rotation symmetry is preserved where the rotation symmetry axis is parallel to the *z*-axis through the H point while the inversion symmetry is broken. In the ZB $C_4N$-II monolayer, the $C_2$ symmetry disappears while the center of inversion symmetry is the H point. Based on their symmetry, we can classify the C-C and C-N bonds. In the ZP $C_3N$ monolayer, all the C atoms are $C_P$ atoms and there are two kinds of $C_P$-$C_P$ bonds. The bond length of $C_P$-$C_P$ is 1.401/1.410 Å. The value of the $C_P$-$C_P$ bond length increases to 1.485/1.493 Å (ZB $C_4N$-I) and 1.491/1.488 Å (ZB $C_4N$-II). In the ZP $C_3N$ monolayer, there are also two kinds of N-$C_P$ bonds, and the bond length is 1.398/1.400 Å. After forming the $C_4N$ monolayer, the length of the N-$C_P$ bond becomes 1.555/1.553 Å (ZB $C_4N$-I) and 1.543/1.544 Å (ZB $C_4N$-II). Different from the ZP $C_3N$ monolayer, there is a $C_R$-$C_P$ bond in the ZB $C_4N$ monolayers, and the bond length is 1.566/1.568 Å (ZB $C_4N$-I) and 1.574/1.586 Å (ZB $C_4N$-II). From the above values of bond length, those in ZB $C_4N$ monolayers are slightly larger than those in ZP $C_3N$ monolayer, because the orbital hybridization of the C/N atoms is changed from ZP $C_3N$ to ZB $C_4N$. All the C/N atoms in the ZP $C_3N$ monolayer are *sp*$^2$ hybridized while all the C/N atoms in ZB $C_4N$ monolayers have *sp*$^3$ hybridization, leading to a buckled structure. Comparing the $C_3N$/$C_4N$ with a hexagonal lattice (h-

C₃N, DB C₄N-I, and DB C₄N-II), the main difference is the six-ring of $C_P$ atoms, which do not exist in the C₃N/C₄N with a rectangular lattice (ZP C₃N, ZB C₄N-I, and ZB C₄N-II) [36]. In the ZB C₄N monolayers, the N atom bonds to three $C_P$ atoms forming three covalent bonds, which can satisfy the eight-electron rule, but the $sp^3$ hybridization of the $C_R$ atom can lead to the appearance of a dangling bond, which is preferred for atomic adsorption [47-50]. Furthermore, the dangling bond of the carbon atom can lead to a special band structure, which will be discussed in the following.

**Table 1.** Structure parameters and cohesive energies ($E_{coh}$, eV/atom) of ZP C₃N and ZB C₄N monolayers. The lattice constants are *a* and *b*, and the $C_P$-$C_P$, N-$C_P$, and $C_R$-$C_P$ represent the bond lengths (Å).

| Monolayer | a | b | $C_P$-$C_P$ | N-$C_P$ | $C_R$-$C_P$ | $E_{coh}$ |
|---|---|---|---|---|---|---|
| ZP C₃N | 4.864 | 4.204 | 1.401/1.410 | 1.398/1.400 | −/− | −7.040 |
| ZB C₄N-I | 4.747 | 4.114 | 1.485/1.493 | 1.555/1.553 | 1.566/1.568 | −6.325 |
| ZB C₄N-II | 4.741 | 4.114 | 1.491/1.488 | 1.543/1.544 | 1.574/1.586 | −6.334 |

Next, we focus on the stability of the three monolayers. Taking ZP C₃N as an example, the cohesive energy is defined as $E_{coh}(C_3N) = [E(C_3N) – 6E(C) – 2E(N)]/8$, where $E(C_3N)$ is the total energy of the ZP C₃N monolayer (per unit cell) and $E(C)/E(N)$ is the total energy of a single C/N atom. The values of cohesive energy are −7.040 eV/atom for ZP C₃N, −6.325 eV/atom for ZB C₄N-I, and −6.334 eV/atom for ZB C₄N-II, respectively. Negative values indicate that these structures are energetically stable. By using $E(C)$ from graphene and $E(N)$ from N₂ molecule, the three values become 0.244 eV/atom, 1.099 eV/atom, and 1.090 eV/atom, respectively, which indicate that the ZP C₃N monolayer is much easier to be achieved experimentally than the two ZB C₄N monolayers, because the C/N atom prefers $sp^2$ hybridization.

Comparing the 0.97 eV/atom of the planar $C_4N$ monolayer [23], the 1.099/1.090 eV atom of the ZB $C_4N$ monolayer means that a special experimental synthesis method should be applied. Fortunately, the similar ZB structure of Si monolayer has been realized experimentally [51], bring a feasible approach for the ZB $C_4N$ monolayers. The h-$C_3N$ monolayer with a hexagonal lattice has a lower $E_{coh}$ (−7.073 eV/atom) than the ZP $C_3N$ monolayer with a rectangular lattice. However, the case is different for the $C_4N$ monolayers. For the $C_4N$-I where the N/$C_R$ atoms are on the same side, the $E_{coh}$ of the ZB $C_4N$-I monolayer is lower than the −6.313 eV/atom of the DB $C_4N$-I monolayer. For the $C_4N$-II where the raised N/$C_R$ atoms are on the opposite side, the $E_{coh}$ of the ZB $C_4N$-II monolayer is almost equal to the −6.334 eV/atom of the DB $C_4N$-II monolayer. Hence, the order of the values of $E_{coh}$ for the four $C_4N$ monolayers is DB $C_4N$-I > ZB $C_4N$-I > ZB $C_4N$-II = DB $C_4N$-II [36]. We also investigated their dynamical stability and found no imaginary frequencies in the calculated phonon spectra of ZP $C_3N$, ZB $C_4N$-I, and ZB $C_4N$-II (Fig. 2), further proving the experimental feasibility of the three monolayers. The corresponding thermal properties are shown in Fig. S1, where the free energy of the monolayer decreases with temperature and the entropy is zero at 0 K, in agreement with the third law of thermodynamics. Furthermore, the thermal stability of the two ZB $C_4N$ monolayers is examined by *ab initio* molecular dynamics (AIMD) simulations with a 2×2×1 supercell at 300 K. The fluctuations of the total energy during 5 ps are shown in Fig. S2, and the corresponding snapshots of the atomic configuration after the AIMD simulations are given, where the ZB structures are well preserved, further confirming their thermal stability.

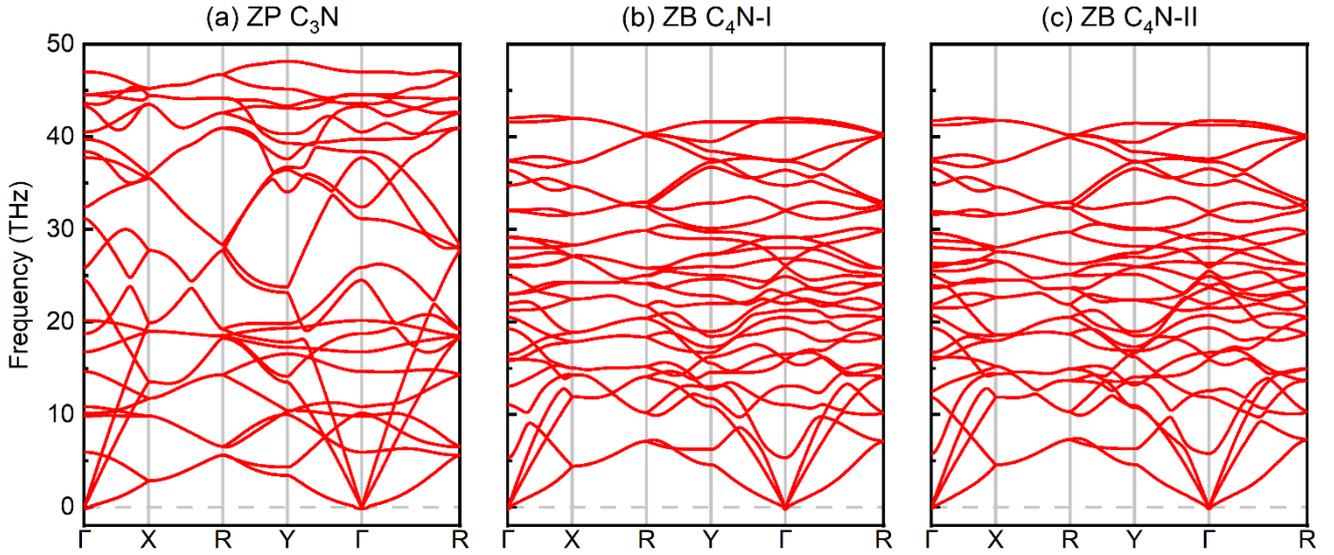

**Fig. 2.** Phonon spectra of ZP $C_3N$ monolayer (a), ZB $C_4N$-I monolayer (b), and ZB $C_4N$-II monolayer (c) along the high-symmetry paths.

## 3.2 Band structure

After the investigation of the structural properties, we calculated the electronic band structures of ZP $C_3N$ monolayer, ZB $C_4N$-I monolayer, and ZB $C_4N$-II monolayer, as shown in Fig. 3. The ZP $C_3N$ monolayer exhibits a metallic band structure, as shown in Fig. 3(b), which is in contrast to the semiconducting band structure with a narrow bandgap (0.39 eV) of h-$C_3N$ monolayer [9]. The ZB $C_4N$-I monolayer and ZB $C_4N$-II monolayer show a semimetal band structure, which is different from the gapless Dirac band structure of DB $C_4N$ monolayer, where the Dirac point is formed by two linear bands at the high-symmetry point K. Since the ZB $C_4N$ monolayers and DB $C_4N$ monolayers have the same number of C/N atoms in the unit cell, it is interesting to question if the ZB $C_4N$ monolayers exhibit a Dirac band structure.

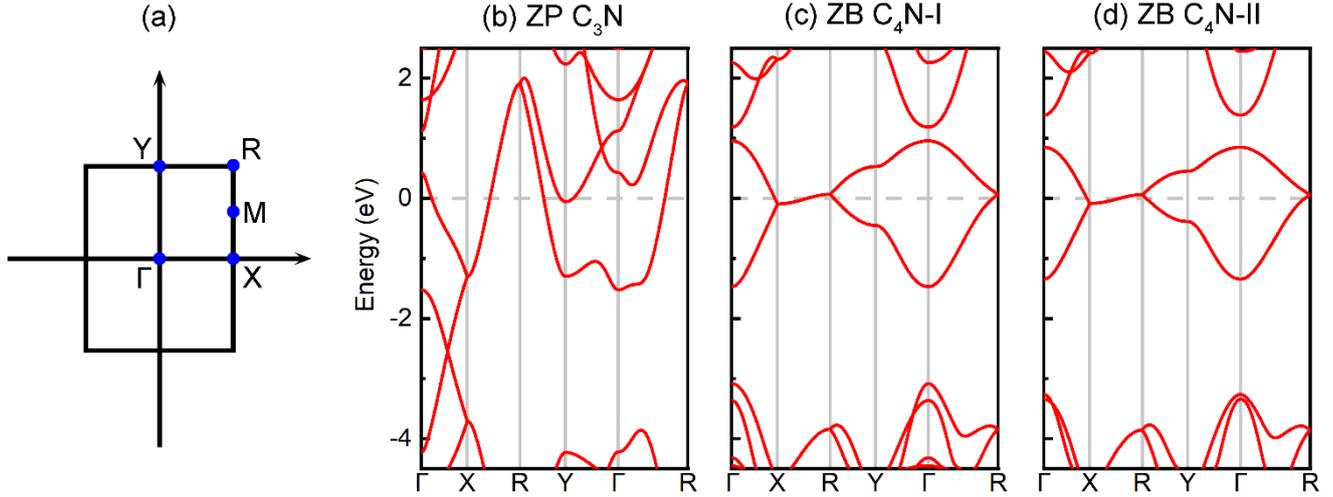

**Fig. 3.** First BZ (a) and electronic band structures for ZP $C_3N$ monolayer (b), ZB $C_4N$-I monolayer (c), and ZB $C_4N$-II monolayer (d). The Fermi level is located at 0 eV.

We will now focus on the band structures of ZB $C_4N$ monolayers. There are two bands around the Fermi level, as shown in Fig. 3(c) and 3(d). It should be noticed that each point of the energy spectrum along the path XR exhibits a zero bandgap, which is double-fold degenerate. In order to study these double-fold degenerate points, we calculated the 2D band structures around the path XR, as shown in Fig. 4(a) (ZB $C_4N$-I) and 5(f) (ZB $C_4N$-II). It is interesting to see that the valence band surface and the conduction band surface meet near the Fermi level forming a slightly curved band along the path XR. To further study this band structure, we chose three special points along the path XR, the initial point (X), the middle point (M), and the end point (R) [Fig. 3(a)], and calculated the band structures in the plane perpendicular to the $k_y$-axis for these three points. Two linear bands form a crossing point near the Fermi level at the point X/M/R, as shown in Fig. 4(b)-5(d) and 4(g)-5(i), and then it can be further concluded that each of the linear crossing points along the path XR should be a Dirac point. However, these Dirac points are different from the Dirac point (cone) at the high symmetry point (path), such as in $C_7N_3/C_{10}N_3/C_{13}N_3/C_{19}N_3$ [18] and $C_5N_4$ [30]. The Dirac points can extend along the $k_y$-axis, showing a 1D behavior. Since it is a slightly curved band along the path XR, which is not an absolute flat band, we call it a quasi-1D Dirac nodal line. A similar band structure was observed in the experiment on bulk NbSi$_{0.45}$Te$_2$ [2], and is essentially different from the ring-shaped nodal line [52], which has been achieved in many 2D systems, including C$_4$N$_4$ [17], C$_9$N$_4$ [34], CuSe [53], H$_{4,4,4}$-graphene [54], Hg$_3$As$_2$ [55], B$_2$C [56], PdS [57], MnN [58], Be$_2$C [59], and Ca$_2$As [60]. These quasi-1D Dirac nodal lines are very rare in 2D systems [61-64]. To our knowledge, the ZB C$_4$N-I/C$_4$N-II is the first predicted monolayer with such quasi-1D Dirac nodal line in the 2D carbon nitrides.

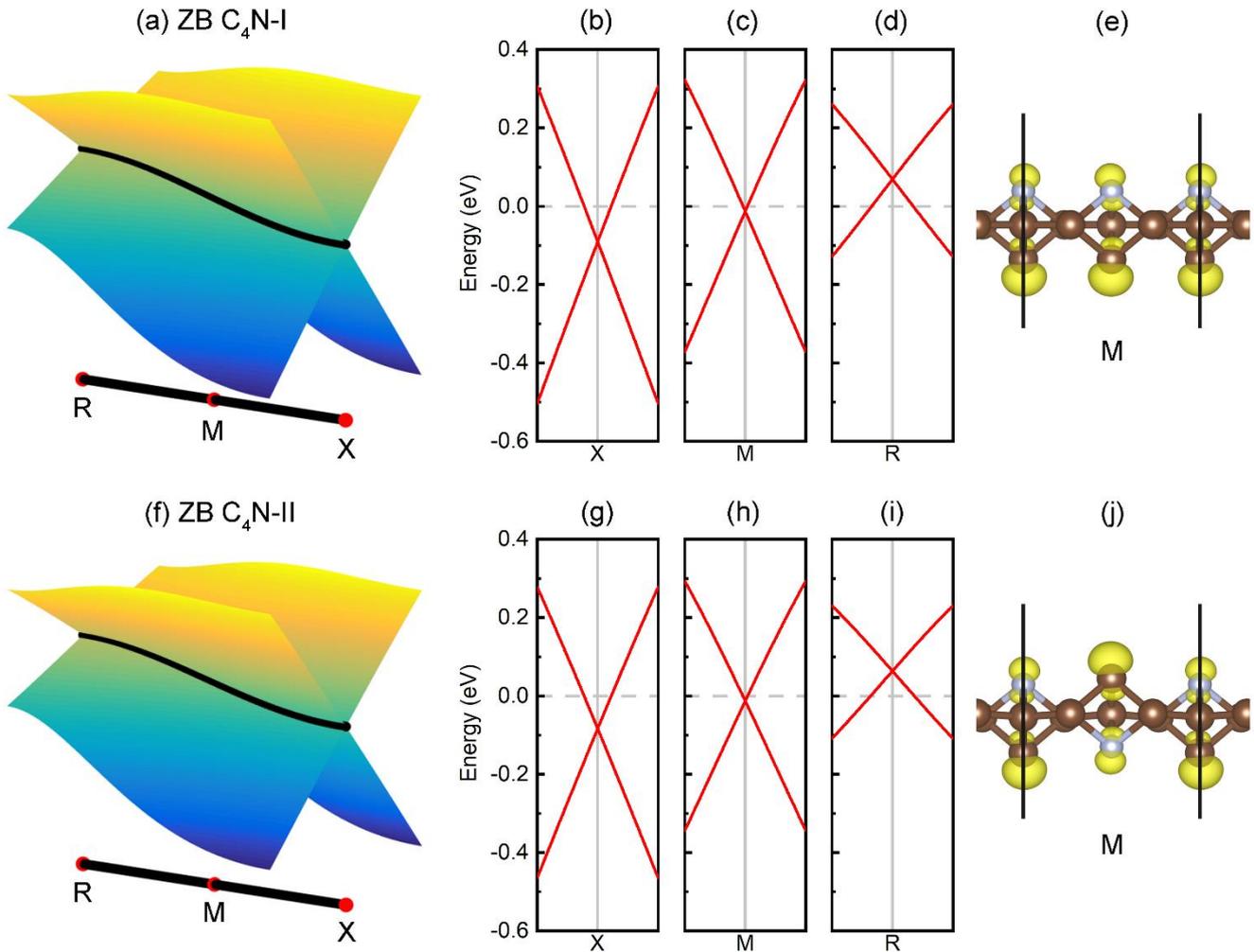

**Fig. 4**. Quasi-1D Dirac nodal lines for ZB C$_4$N-I monolayer (a) and ZB C$_4$N-II monolayer (f). Band structures in the plane perpendicular to the $k_y$-axis for the three points (X, M, and R) of ZB C$_4$N-I monolayer (b)-(d) and ZB C$_4$N-II monolayer (g)-(i). The corresponding Kohn-Sham wave functions of the Dirac point

(point M) for ZB C$_4$N-I monolayer (e) and ZB C$_4$N-II monolayer (j). The isosurface values were set to 0.02 eÅ$^{-3}$.

Since the Dirac point is formed by two linear band lines, the Fermi velocity is an important parameter. The Fermi velocities are $3.7\times10^5$ m/s (X), $3.3\times10^5$ m/s (M), and $1.8\times10^5$ m/s (R) for the ZB C$_4$N-I monolayer, while the Fermi velocities are $3.4\times10^5$ m/s (X), $3.0\times10^5$ m/s (M), and $1.6\times10^5$ m/s (R) for the ZB C$_4$N-II monolayer. The numerical variation of the Fermi velocity is shown in Fig. 5(a), and a clear downward trend from X to R can be seen, where the values of ZB C$_4$N-I monolayer are $1.8\sim3.7\times10^5$ m/s while the values of ZB C$_4$N-I monolayer are $1.6\sim3.4\times10^5$ m/s, which are comparable to that of DB C$_4$N monolayer [36]. For each corresponding Dirac point along the path XR, the Fermi velocity of ZB C$_4$N-I monolayer is slightly larger than that of ZB C$_4$N-II monolayer. The quasi-1D Dirac nodal line is composed of a number of Dirac points with varying Fermi velocity, further confirming the quasi-1D behavior. The ZB C$_4$N-I monolayer is protected by time-reversal symmetry $T$ and the glide mirror symmetry $M_y$ parallel to the xz-plane, which satisfies $M_y^2 = e^{-ik_x a}$. Due to $[T, M_y] = 0$ and $T^2 = 1$, the $(TM_y)^2 = e^{-ik_x a}$ can be obtained. For the path XR, each point should be $k_x a = \pi$, and is invariant under the antiunitary symmetry $TM_y$, $(TM_y)^2 = -1$. Therefore, the bands along the path XR should be doubly degenerate, leading to the quasi-1D nodal line [65]. For the ZB C$_4$N-II monolayer, it becomes the screw axis symmetry along the x-direction ($C_{2x}^2 = e^{-ik_x a}$) [66], and the case is similar to that of the ZB C$_4$N-I monolayer, also leading to the quasi-1D nodal line. Considering the spin-orbit coupling (SOC), each Dirac point can open a bandgap, and the values are 0.46~0.65 meV for the ZB C$_4$N-I (0.06~0.20 meV for the ZB C$_4$N-II), which can be neglected due to the light elements C and N. The corresponding edge states are shown in Fig. S3, where no edge states can be seen, indicating the quasi-1D nodal lines are topological trivial.

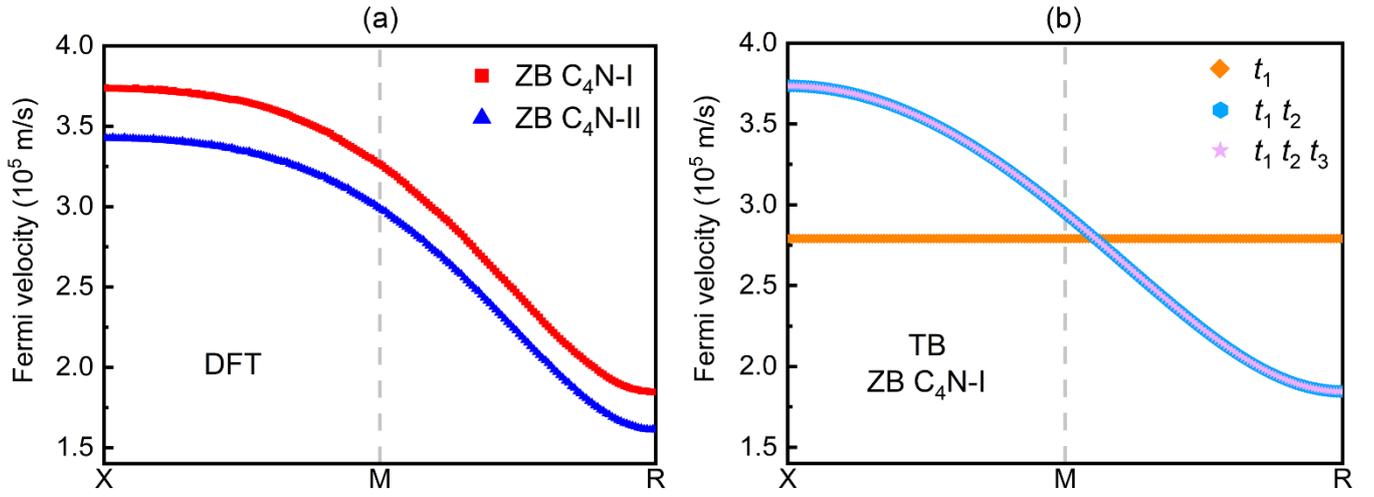

**Fig. 5.** (a) Numerical variation of Fermi velocity along the path XR for ZB $C_4N$-I monolayer (red) and ZB $C_4N$-II monolayer (blue). The results are from DFT. (b) Numerical variation of Fermi velocity along the path XR for ZB $C_4N$-I monolayer. The results are from TB.

In order to analyze the relationship between band structure and atomic orbitals, we calculated the projected band structures for different atomic orbitals, as shown in Fig. 6(a)-6(d) (ZB $C_4N$-I, red) and 7(e)-7(h) (ZB $C_4N$-II, blue). It is clear that the two bands close to the Fermi level are mainly from the $p_z$ orbitals of the C and N atoms, and the $s/p_x/p_y$ atomic orbitals of the C and N atoms have very small contributions to those bands. However, not all the C atoms are important. As illustrated in Fig. 4(e) and 4(j), only the $C_R$ and N atoms contribute to the formation of the Dirac point at the point M, and the corresponding Kohn-Sham wave functions around the $C_R$ atoms are larger than those around the N atoms. Hence, we can conclude that only the $p_z$ atomic orbital of the $C_R$ atom plays an important role in forming the band structure around the Fermi level. By using this conclusion, the similar band structures of the two ZB $C_4N$ monolayers can be totally understood. The $C_R$ atoms in the ZB $C_4N$-I can form a planar $C_R$ monolayer, where there is no buckled height for the two $C_R$ atoms in the unit cell, while those in the ZB $C_4N$-II can form a buckled

$C_R$ monolayer, where there is a buckled height for the two $C_R$ atoms in the unit cell. The planar and buckled $C_R$ monolayers can be in analogy with graphene and silicene (germanene). Since the contributions around the Fermi level are mainly from the $p_z$ atomic orbitals of the $C_R$ atom, the buckled height is not important for the formation of the band structure. Hence in the following, we will only focus on the planar $C_R$ monolayer from the ZB $C_4N$-I.

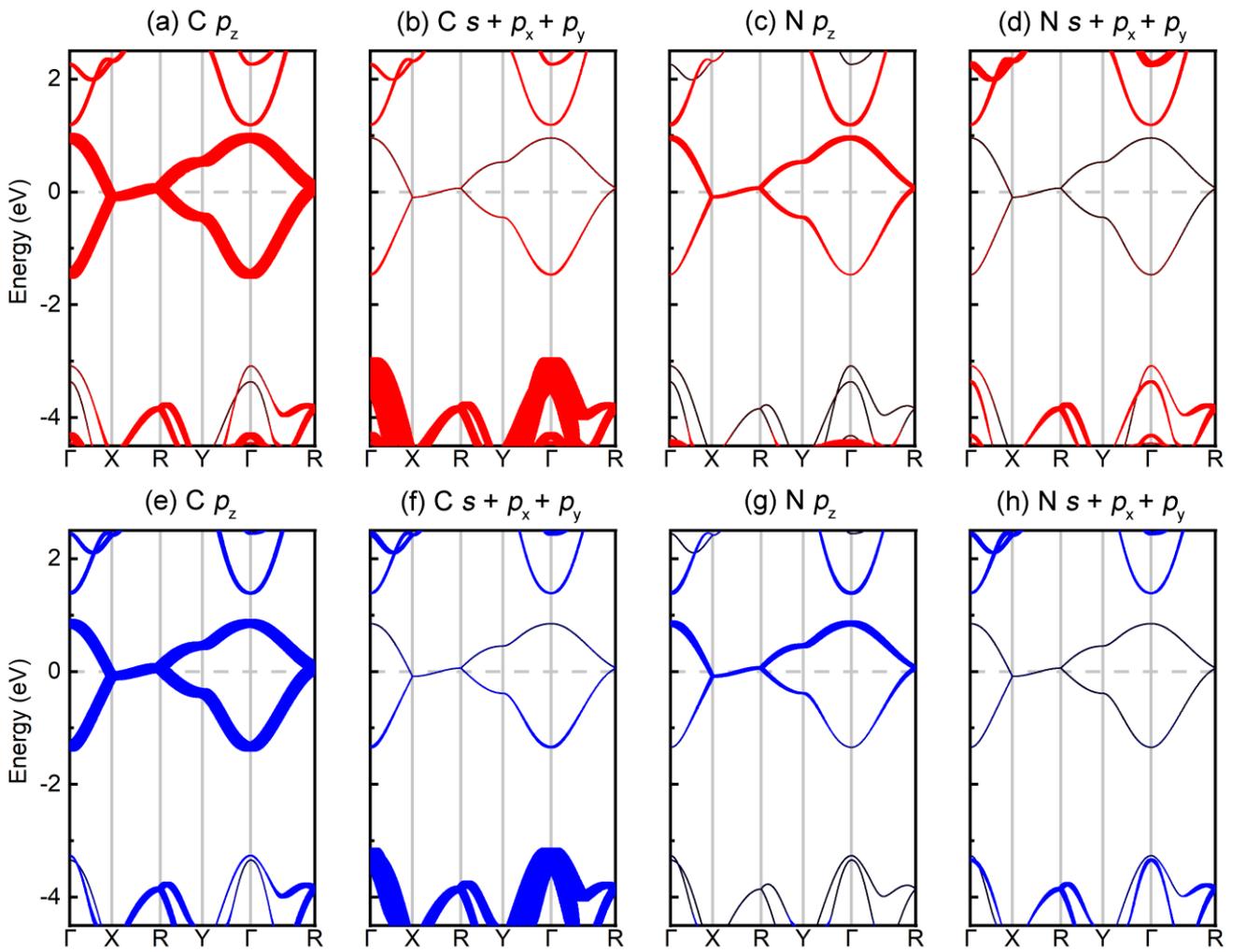

**Fig. 6.** Projected band structures of ZB $C_4N$-I monolayer (a)-(d) and ZB $C_4N$-II monolayer (e)-(h). The thickness of the curve is a measure of the contribution of the different atomic orbitals.

### 3.3 Tight-binding and Su-Schrieffer-Heeger models

The quasi-1D Dirac nodal line originates from the corresponding quasi-1D structure, which can be further understood from a tight-binding (TB) model. The ZB C$_4$N-I monolayer can be simplified as a planar C$_R$ monolayer where the two C$_R$ atoms in the unit cell are at the same height of the z-axis, as shown in Fig. 7(a). Based on the simplified structure, a TB model with $p_z$ atomic orbital can be constructed. The on-site energy of the two C$_R$ atoms in the unit cell can be set to zero due to their equality. There are three hopping parameters, which are labeled as $t_1$, $t_2$, and $t_3$, corresponding to the first, second, and third nearest-neighbor (NN) interaction [67], respectively, as shown in Fig. 7(a). The $t_1$, $t_2$, and $t_3$ can satisfy the glide mirror symmetry $M_y$. The absolute value of the hopping parameter is related to the distance between the two C$_R$ atoms. The larger the distance, the smaller its absolute value. The C$_R$-C$_R$ distance is 2.780/3.570/4.114 Å, corresponding to $t_1$/$t_2$/$t_3$ ($t_1 < t_2 < t_3 < 0$). First, we consider $t_1$ without $t_2$ and $t_3$. The Hamiltonian with only the first NN interaction can be written as

$$H_{first} = t_1 \sum_i (a_{i,j}^+ b_{i,j} + a_{i,j+1}^+ b_{i,j} + h.c.),$$

where $a_{i,j}$ and $b_{i,j}$ are the particle operators for A and B sites in a unit cell. After Fourier transform to momentum space, the corresponding kernel of the Hamiltonian is

$$H_{first}(\vec{k}) = \begin{bmatrix} 0 & t_1 e^{-ik_y b'}(e^{i\frac{k_x a}{2}} + e^{-i\frac{k_x a}{2}}) \\ t_1 e^{ik_y b'}(e^{-i\frac{k_x a}{2}} + e^{i\frac{k_x a}{2}}) & 0 \end{bmatrix},$$

where $b' < b$ [Fig. 7(a)]. The energy spectrum is then given by $E = \pm 2|t_1 \cos(\frac{k_x a}{2})|$. The gapless Dirac point can be found at each point along the path XR ($k_x a = \pi$), and extends along the $k_y$-axis. The Fermi velocity along the path XR is a constant, $v_F = \frac{a}{\hbar}|t_1|$. Next, we consider $t_1$ and $t_2$ without $t_3$. The Hamiltonian describing the second NN interaction can be written as

$$H_{second} = t_2 \sum_i (a^+_{i-1,j} b_{i,j} + a^+_{i-1,j+1} b_{i,j} + h.c.).$$

After Fourier transform to momentum space, the corresponding kernel of the Hamiltonian is

$$H_{second}(\vec{k}) = \begin{bmatrix} 0 & t_2 e^{ik_y(b-b')}(e^{i\frac{k_x a}{2}} + e^{-i\frac{k_x a}{2}}) \\ t_2 e^{-ik_y(b-b')}(e^{-i\frac{k_x a}{2}} + e^{i\frac{k_x a}{2}}) & 0 \end{bmatrix}.$$

The calculated relationship between $E$ and $\vec{k}$ is now $E = \pm 2|\sqrt{t_1^2 + t_2^2 + 2t_1 t_2 \cos k_y b} \cos(\frac{k_x a}{2})|$. The gapless Dirac point is found at each point along the path XR ($k_x a = \pi$), and extends along the $k_y$-axis. The Fermi velocity along the path XR is now $v_F = \frac{a}{\hbar}\sqrt{t_1^2 + t_2^2 + 2t_1 t_2 \cos k_y b}$, varying along $k_y$-axis [3,68]. The maximum value ($v_F = \frac{a}{\hbar}|t_1 + t_2|$) is at the point X ($k_y b = 0$) while the minimum value ($v_F = \frac{a}{\hbar}|t_1 - t_2|$) is at the point R ($k_y b = \pi$). Finally, we consider $t_1$, $t_2$, and $t_3$. The Hamiltonian describing the third NN interaction can be written as

$$H_{thrid} = t_3 \sum_i (a^+_{i-1,j} a_{i,j} + a^+_{i+1,j} a_{i,j} + b^+_{i-1,j} b_{i,j} + b^+_{i+1,j} b_{i,j}).$$

After Fourier transform to momentum space, the corresponding kernel of the Hamiltonian is

$$H_{third} = \begin{bmatrix} t_3(e^{ik_y b} + e^{-ik_y b}) & 0 \\ 0 & t_3(e^{ik_y b} + e^{-ik_y b}) \end{bmatrix}.$$

The energy spectrum is $E = \pm 2|\sqrt{t_1^2 + t_2^2 + 2t_1 t_2 \cos k_y b} \cos(\frac{k_x a}{2})| + 2t_3 \cos(k_y b)$. The gapless Dirac point can be found at each point along the path XR ($k_x a = \pi$), and extends along the $k_y$-axis. The Fermi velocity along the path XR should be $v_F = \frac{a}{\hbar}\sqrt{t_1^2 + t_2^2 + 2t_1 t_2 \cos k_y b}$, which is the same as in previous case including $t_1$ and $t_2$ without $t_3$. The $v_F = \frac{a}{\hbar}|t_1 + t_2|$ ($k_y b = 0$) corresponds to the Fermi velocity (X) from

DFT while the $v_F = \frac{a}{\hbar}|t_1 - t_2|$ ($k_y b = \pi$) corresponds to the Fermi velocity (R) from DFT. The values of $t_1$ and $t_2$ were calculated to be −0.387 eV and −0.131 eV, respectively. The energy difference of the maximum value (R) and minimum value (X) along the path XR should be $4|t_3|$, and then the $t_3$ is −0.040 eV. The TB band structure (blue) including the $t_1$, $t_2$, and $t_3$ is in good agreement with our DFT result (red), as shown in Fig. 7(d). Furthermore, the Fermi velocities from the TB model can also be in good agreement with the DFT result, as shown in Fig. 5(b). For the quasi-1D Dirac nodal line of the $C_4N$-I monolayer as obtained from DFT, there are three important features along the path XR: (i) the gapless Dirac points, (ii) varying Fermi velocity, and (iii) slightly curved band which is not a totally flat band. It can be seen that the features (i)/(ii)/(iii) originate from the $t_1/t_2/t_3$, as shown in Fig. 7(b)/8(c)/8(d). The reliability of above TB model was further confirmed in Fig. S4, where the cases of strain and defect were considered and the results form TB also are in good agreement with those from DFT.

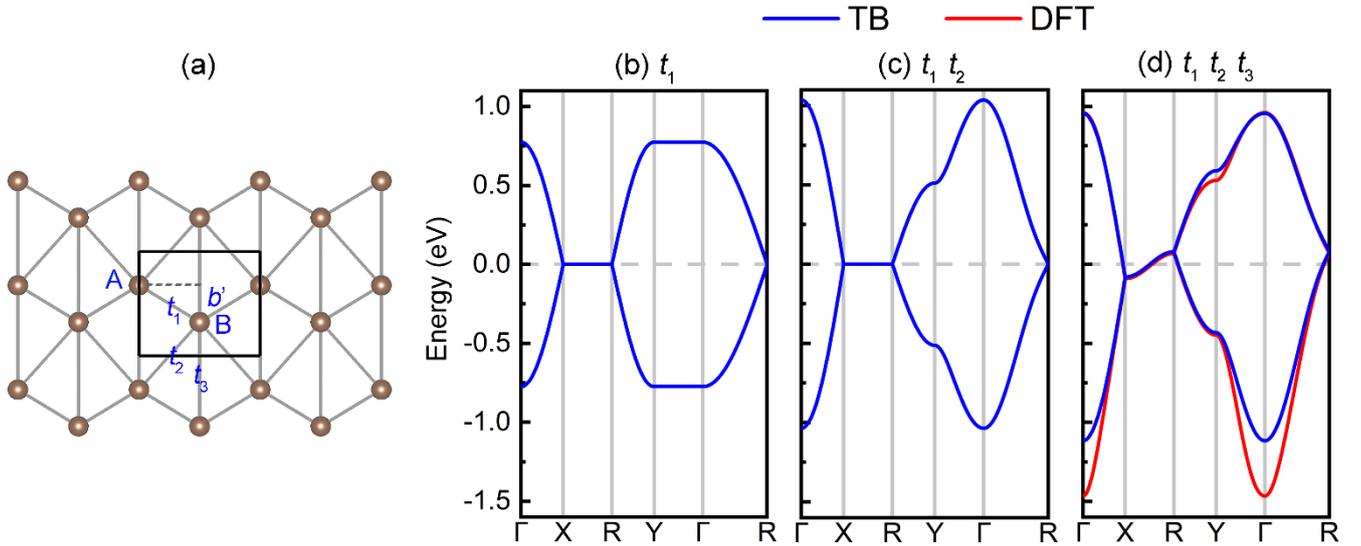

**Fig. 7.** (a) Simplified $C_R$ monolayer and hopping parameters for TB model. TB band structures (blue) when considering the NN interactions up to the (b) first ($t_1$), (c) second ($t_1$ and $t_2$), and (d) third ($t_1$, $t_2$, and $t_3$). (d) DFT band structure (red).

The above TB results can be understood by the simplified $C_R$ monolayer in Fig. 7(a), which is composed of zigzag $C_R$ chains. Since the interaction between the two nearest-neighbor $C_R$ atoms in the same chain is much larger than that between the different zigzag $C_R$ chains, the final band structure along the path XR is similar to that of a single $C_R$ chain, which is a 1D structure. By setting $t_2 = 0$ and $t_3 = 0$, there is no interaction between the zigzag $C_R$ chains, and therefore an absolute flat band along the path XR is obtained, where the 1D structure corresponds to the 1D Dirac nodal line. By setting $t_2 = -0.131$ eV and $t_3 = -0.040$ eV, interactions between the zigzag $C_R$ chains are introduced, and therefore a slightly curved band with varying Fermi velocity along the path XR is obtained, where the quasi-1D structure corresponds to the quasi-1D Dirac nodal line. In NbSi$_x$Te$_2$ [3], which has an almost flat band along the high-symmetry path, the third NN interaction can be neglected. Here, the main interaction between the $C_R$ atoms in the zigzag chain leads to Dirac points originating from the first NN ($t_1$), while the weak interactions between the zigzag $C_R$ chains result in the varying Fermi velocity originating from the second NN ($t_2$) and the slightly curved band structure originates from the third NN ($t_3$).

Comparing the edge states from DFT+Wannier [Fig. S3(a)] [69-72], our TB model for the nanoribbon with width ≈ 20nm also gives similar topological trivial edge states, as shown in Fig. S5(c). The spectrum is $E = \pm 2 |\sqrt{t_1^2 + t_2^2 + 2t_1 t_2 \cos k_y b} \cos(\frac{k_x a}{2})| + 2t_3 \cos(k_y b)$, and the term of $\sqrt{t_1^2 + t_2^2 + 2t_1 t_2 \cos k_y b}$ can play an important role in determining the topological properties according to the Su-Schrieffer-Heeger (SSH) model [73], where the winding number is 0 for the $t_1 = -0.387$ eV and $t_2 = -0.131$ eV [Fig. S5(a) and S5(b)]. Moving the B site along the y-axis [Fig. S5(a) and S5(d)], the value of $|t_1|$ will decrease while the value of $|t_2|$ will increase. When $t_1$ is equal to $t_2$ ($t_1 = t_2 = -0.259$ eV), another gapless nodal line can be formed along the path RY ($k_y b = \pi$) [Fig. S5(e)], because the term of $\sqrt{t_1^2 + t_2^2 + 2t_1 t_2 \cos k_y b}$ becomes 0.

Due to the nodal line along X-R-Y, the band structure becomes a ring-shaped nodal line in the first BZ, which is the critical point of a topological phase transition. The corresponding edge states are shown in Fig. S5(f), which are indistinguishable. When we keep moving the B site along the *y*-axis [Fig. S5(g)], the values of $t_1$ and $t_2$ can be inversed ($t_1 = -0.131$ eV and $t_2 = -0.387$ eV) where the winding number becomes 1, and the bandgap can be reopened along the path RY [Fig. S5(h)]. The corresponding edge states are shown in Fig. S5(i), where the obvious topological edge states can be seen. Similar to the SSH model [74], the topological properties of our system along the *y*-axis depend on the relative value of $t_1$ and $t_2$, and the critical state corresponds to the gapless nodal line along the path RY. Although the quasi-1D nodal line along path XR exhibits the topological trivial property, the band structure along the path RY can give a phase transition. When only considering the path GY, $E = \pm 2|\sqrt{t_1^2 + t_2^2 + 2t_1 t_2 \cos k_y b}| + 2t_3 \cos(k_y b)$, the standard 1D SSH model can be obtained, the topological properties are characterized by the so-called Zak phase. The Zak phase of the *n*-th band $\gamma_n$ is integral over the BZ of the Berry connection:

$$\gamma_n = i \int_{-\pi/b}^{\pi/b} \langle u_{nk} | \partial_k | u_{nk} \rangle dk,$$

where $|u_{nk}\rangle$ is the periodic part of the Bloch function of *n*-th band with momentum *k*. The total Zak phase, $\varphi_{Zak}$, is the sum of $\gamma_n$ over the occupied bands, where the only one occupied band needs to be considered. For the Figure S5(a), $\varphi_{Zak} = \pi$, which is topological nontrivial phase, while for the Figure S5(d), $\varphi_{Zak} = 0$, which is topological trivial phase.

### 3.4 Topological property

Although we have confirmed origin of the topological phase transition by the above TB&SSH model, a natural question arises: How to realize the topological nontrivial phase in the real material system? Here, we take the C$_4$N-I monolayer as an example. According to the SSH model, the key is keeping different boundaries, similar to the SSH model. As shown in Fig. 8, there are two phases for the C$_4$N-I monolayer, including the strong interaction (between the two C$_R$ atoms, red solid line) in the unit cell and the weak interaction (between the two C$_R$ atoms, red dotted line) in the unit cell [75]. Although the phases actually exhibit the same structure, there is a parity reversal at the high-symmetry point Y [76], indicating a topological phase transition. When cutting the edge of the C$_4$N-I monolayer with a weak interaction (red dotted line), the trivial edge states can be obtained, as shown in Fig. 8(a). However, when cutting the edge of the C$_4$N-I monolayer with a strong interaction (solid line), the nontrivial edge states can be obtained, where a flat band can be seen around the Fermi level, as shown in Fig. 8(b). Here, the bulk gap is determined by the gap of the point Y, and the band inversion at the point Y results in the topological nontrivial edge sates. Therefore, we obtain the SSH edge states in the 2D C$_4$N-I system.

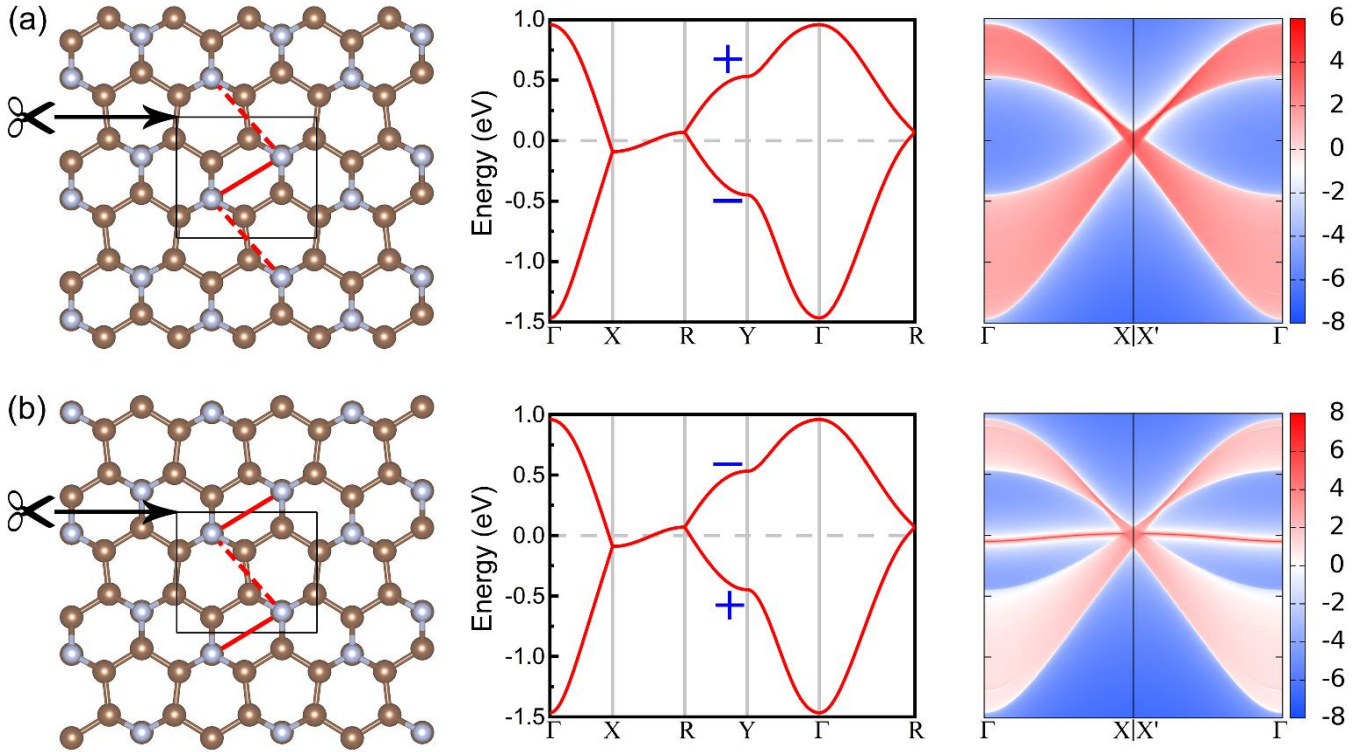

**Fig. 8.** Two phases for the C$_4$N-I monolayer. (a) is the trivial phase, and (b) is the nontrivial phase.

## 4. CONCLUSION

In summary, we propose two new ZB $C_4N$ monolayers, which are a buckled structure and have a quasi-1D Dirac nodal line. Different from most experimental and theoretical monolayers of carbon nitride that include the edge-N and graphitic-N atoms, the raised-N atoms with $sp^3$ hybridization are important in forming our ZB $C_4N$ monolayers, which can lead to the appearance of dangling bonds of the carbon atoms. The semimetal band structure of the quasi-1D Dirac nodal line is a highlight for the ZB $C_4N$ monolayers. Our TB model considering the first, second, third NN interactions for the zigzag carbon chains not only is able to reproduce the DFT band structure, but also shows the correspondence between the features of quasi-1D Dirac nodal line and the NN interactions, which offers a good platform for further research of quasi-1D physics. Although the quasi-1D Dirac nodal line exhibits the topological trivial property, the topological nontrivial phase can be achieved by changing the selection of unit cell, which is a boundary-dependent topological flat edge states determined by the determined by the Zak phase and band inversion. Our calculated results from DFT and TB provides a new light on fundamental understanding of the SSH model in 2D materials.

**Acknowledgments**

We thank Xiangru Kong and Adrien Bouhon for helpful discussions. This work is supported by the National Natural Science Foundation of China (Grant No. 12004097). F.M.P. acknowledges financial support from the HSE University Basic Research Program. X.L. acknowledges financial support from the National Natural Science Foundation of China (Grant No. 22005087). The computational resources utilized in this research were provided by Shanghai Supercomputer Center.